\documentclass[aps,prl,twocolumn,superscriptaddress]{revtex4}

\usepackage{amsfonts,amssymb,amsmath,latexsym,epsfig,wasysym} 
\usepackage[sort&compress]{natbib}

\newcommand{\tcoll}{\langle \tau \rangle}
\newcommand{\ep}{\langle R e^{\kappa \tau} \rangle_c}

\begin{document}

\title{Chaotic Systems with Absorption}

\author{Eduardo G. Altmann\footnote{Equal contribution of the three authors}}
\affiliation{Max Planck Institute for the Physics of Complex Systems, 01187 Dresden, Germany}
\author{Jefferson S. E. Portela}
\affiliation{Fraunhofer Institute for Industrial Mathematics ITWM, 67663 Kaiserslautern, Germany}
\author{Tam\'as T\'el}
\affiliation{Institute for Theoretical Physics - HAS Research Group, E\"otv\"os University, Budapest, H--1117, Hungary}

\begin{abstract}
Motivated by applications in optics and acoustics we develop a dynamical-system approach to describe absorption in chaotic systems. We
introduce an operator formalism from which we obtain (i) a general formula  for the escape rate~$\kappa$ in terms of the natural
conditionally-invariant measure of the system, (ii) an increased multifractality when 
compared to the spectrum of dimensions~$D_q$ obtained without taking absorption 
and return times into account, and (iii) a generalization of the Kantz-Grassberger formula that expresses $D_1$ in
terms of $\kappa$, the positive Lyapunov exponent, the average return time, and a new quantity, the reflection rate. Simulations in the
cardioid billiard confirm these results. 
\begin{center}{\bf Published as: Phys. Rev. Lett. 111, 144101 (2013)}\end{center}
\end{abstract} 


\maketitle


The design of concert halls was probably the first problem in which the importance of the partial absorption of energy along trajectories was
fully recognized~\cite{Kuttruff:2005,Joyce:1975}. In Berry's elegant formulation, confinement is needed 
{\it 
to prevent sound from being attenuated by radiating into the open air. But if the confinement were perfect, that is, if the walls of the
room were completely reflecting, sounds would reverberate forever. To avoid these extremes, the walls in a real room must be partially
absorbing}~\cite{BerryFor}.
Besides acoustics~\cite{Kuttruff:2005,Joyce:1975,BerryFor,Mortessagne:1992,Mortessagne1993,Tanner:1998,Tanner:2013}, chaotic dynamical systems in which trajectories are partially absorbed appear nowadays in an increasing number
of different areas~\cite{RMP}, ranging from optics (microlasers)~\cite{Harayama2011,Wiersig2008} to environmental sciences (resetting
mechanism)~\cite{Pierrehumbert2007} and quantum chaos~\cite{Nonnenmacher:2008}. 
The analogy of the decay of the sound intensity with the survival probability 
of transient chaos has early been recognized~\cite{Mortessagne1993},
here we add that a sharp distinction between the attenuation of energy (absorption) and the escape of particles (transport) is necessary.

\begin{figure}[!bt]
\includegraphics[width=1\columnwidth]{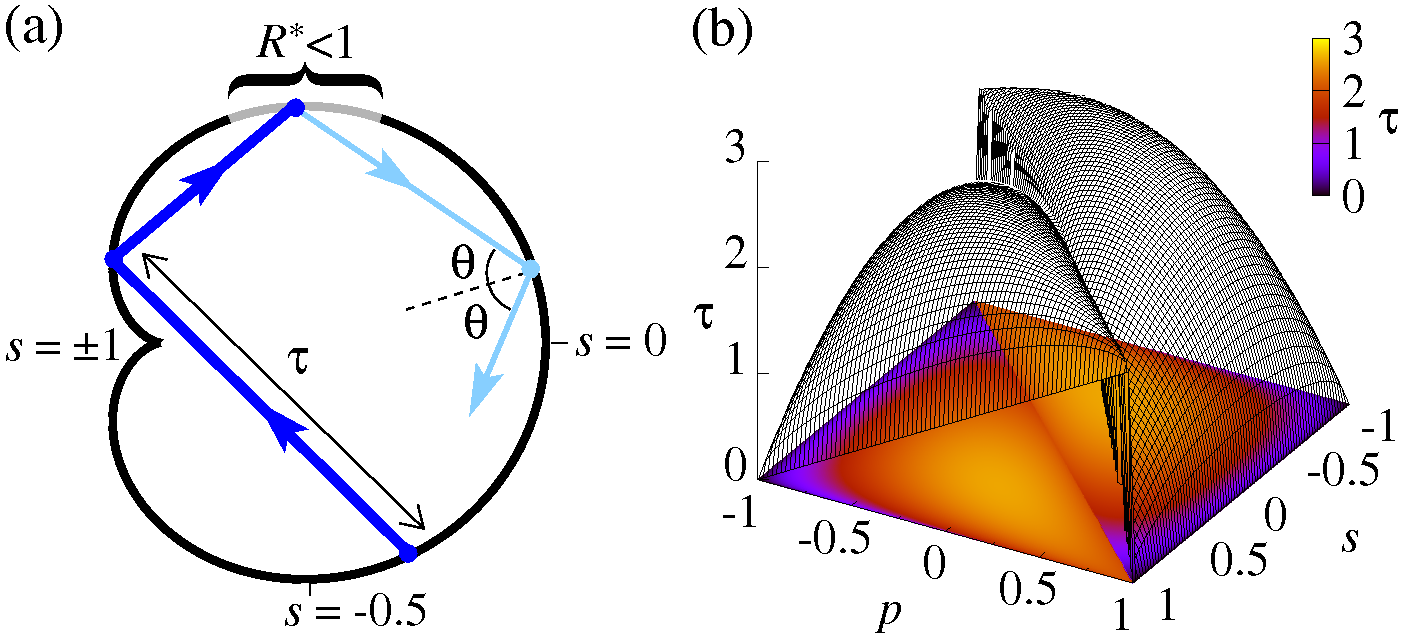}
\caption{Billiards naturally incorporate both partial reflection at the boundary and non-trivial return times between collisions. (a)
  Cardioid billiard, whose boundary in polar coordinates is $r(\phi) = 1 + \cos(\phi)$~\cite{Robnik1983}. 
  The intensity $J$ of the rays decays due to $R(\vec{x})=R^*<1$ in the gray boundary interval at the top ($R(\vec{x})=1$ otherwise). (b) Return time
  distribution $\tau(\vec{x})$ in the cardioid billiard (velocity modulus is chosen to be unity). Birkhoff coordinates $\vec{x}=(s,p=\sin{\theta})$ are used where $s$ is the
arc length along the boundary and $\theta$ is the collision angle.}
\label{fig1}
\end{figure}

A seemingly unrelated  problem is monitoring continuous time in flows represented by discrete-time
maps $ \vec{x}_{n+1} = f(\vec{x}_n)$ through a proper Poincar\'e surface of section. 
Both problems can be handled extending the phase space of map $f$~\cite{Gas-book} to include the true physical time $t_n$ and the ray intensity $J_n$ at the $n$-th
intersection with the Poincar\'e section as
\begin{equation}\label{eq.extended}
f_{\text{extended}}:\left\{\begin{array}{lll}
\vec{x}_{n+1} & =f(\vec{x}_{n}),  \\
t_{n+1} & = t_n + \tau(\vec{x}_{n}),\\
J_{n+1} & = J_n R(\vec{x}_{n}),
\end{array}
\right.
\end{equation}
where the return time $\tau(\vec{x})\ge0$, chosen as the time between intersections $\vec{x}$
  and $\vec{x}'\equiv f(\vec{x})$, and the reflection coefficient
$0<R(\vec{x})\le 1$ are functions of the coordinate $\vec{x}$ on the Poincar\'e section.
Probably the most prominent systems incorporating both properties
are billiards such as the one in Fig.~\ref{fig1}. Concert halls can be modeled as $3$D billiards~\cite{Kuttruff:2005}.    
{\it Trajectory-based} simulations (ray tracing~\cite{Harayama2011,Tanner:1998}) in these systems are performed from Eq.~(\ref{eq.extended}) by tracking $t$ and $J$ along each trajectory.

In this Letter, we show that absorption and true time 
lead to surprising modifications of fundamental results of chaotic dynamics. 
This is done by introducing an {\it operator-based} formalism. We use it to derive an expression for the escape
rate~$\kappa$ as a function of the natural  {\it conditionally-invariant} measure of the system.
As a consequence, we show that $\kappa$  depends on the entire distributions of $\tau(\vec{x})$ and $R(\vec{x})$ and not only on their averages.
In terms of the spectrum of fractal dimensions $D_q$ of the invariant sets, we show that $\tau$ and $R$ typically enhance multifractality and
that $D_1$ can be expressed as a function of $\kappa$, the average Lyapunov exponent, and a new parameter. 

We start with the well-known operator formalism for open maps~\cite{Gas-book,PY:1979,Tel:1987,LaiTel-book}.
The escape rate $\kappa$ of an open (possibly non-invertible) map ${f}$ is related to the largest eigenvalue ${e^{-\kappa}}$ of the Perron-Frobenius operator acting on the
density of trajectories $\varrho$ 
\begin{equation}
 \varrho_{n+1}(\vec{x}') = \sum_{\vec{x}\in f^{-1}({\vec{x}')}} {e^{\kappa}} 
                \frac{\varrho_{n}({\vec{x})}}{\mid \mathcal{D}_f(\vec{x})\mid}, 
\label{eq.FP1}
\end{equation}
where $\mathcal{D}_f(\vec{x})$ is the Jacobian at $\vec{x}$. 
Equation (\ref{eq.FP1}) expresses that the probability in a small region around $\vec{x}$ at step $n$ is the same as the $f$-image of that region at step $n+1$, when compensating for the escape. 
 $\kappa$ follows from the
requirement that the integral of $\varrho_n$ over a fixed phase space region
containing the underlying nonattracting chaotic set (a repeller or a saddle) remains finite for $n \rightarrow \infty$.
In this limit, $\varrho_n \rightarrow \varrho_c$ concentrates on the unstable manifold of the chaotic saddle according to
the conditionally invariant measure (c-measure)~\cite{PY:1979}.

We now introduce an operator formalism for the extended map~(\ref{eq.extended}).  Imposing a uniform decay of trajectories in time $t$, instead of the number $n$ of iterations,
it is natural to replace~$e^\kappa$ in Eq.~(\ref{eq.FP1}) by $e^{\kappa \tau(\vec{x})}$.  
The reflection coefficient $R(\vec{x})$ corresponds to an immediate loss of intensity and is therefore
introduced also on the right hand side of Eq.~(\ref{eq.FP1}). Altogether, the density
function $\rho$ of map~(\ref{eq.extended}) evolve as
\begin{equation}
 \rho_{n+1}(\vec{x}') =  \sum_{\vec{x}\in {f}^{-1}({\vec{x}')}}  e^{\kappa \tau(\vec{x})}
                \frac{R(\vec{x}) \rho_{n}(\vec{x})}{\mid \mathcal{D}_f(\vec{x})\mid}. 
\label{eq.FP2}
\end{equation}
This operator  generalizes the true-time formalism of Gaspard~\cite{Gaspard:1996}
and Kaufmann \& Lustfeld~\cite{Kaufmann2001} by introducing reflection in a similar spirit as in Tanner's work on driven acoustic systems~\cite{Tanner:1998,Tanner:2013}.  
Among the different generalized transfer operators~\cite{Faure} and other possible
  generalizations of Eq.~(\ref{eq.FP1}),  Eq.~(\ref{eq.FP2}) is the one that remains
  faithful to the physical picture used in the extension of maps~$f$ to extended maps
  $f_{\text{extended}}$ in Eq.~(\ref{eq.extended}). Indeed, the operator we recently
  introduced~\cite{RMP} differs from Eq. (\ref{eq.FP2}) precisely because of the
  different convention of $\tau$ (defined  as a function of the endpoint $\vec{x}'$) in
  Eq.~(\ref{eq.extended}). Equation~(\ref{eq.FP2}) is an extension to non-invertible maps
  of this previously  defined operator.
For $n\rightarrow\infty$, $\rho_0$ approaches
a limit distribution $\rho_{\infty}$ (of finite integral) which is $\rho_c$ associated to the c-measure~$\mu_c$ of the extended map
(\ref{eq.extended}), normalized over the region of interest, $\Omega$, on the Poincar\'e
map.  The support of $\rho_c$ and $\varrho_c$ from~(\ref{eq.FP1}) coincide, but the
densities are typically different. 
In open systems there is a {\it region of escape} $E\subset\Omega$ in which trajectories
escape $\Omega$ within one iteration of the Poincar\'e
  map~$f$. Because this escape is not due to absorption and happens instantaneously, we choose $R(\vec{x})=1$
  and $\tau(\vec{x})=0$ for $\vec{x}\in{E}$. 

We can now derive a relation for~$\kappa$ as a function of $\rho_c$. By integrating, for $n \rightarrow \infty$, both sides of Eq.~(\ref{eq.FP2}) over $\Omega$ we obtain 
\begin{equation}\label{eq.escaperate'}
\begin{array}{lllll}
1
     & = \int_\Omega d\vec{x}' e^{\kappa \tau(\vec{x})}  \dfrac{R(\vec{x}) \rho_{c}(\vec{x})}{\mid \mathcal{D}_f(\vec{x})\mid}_{\vert_{\vec{x} = f^{-1}(\vec{x}')}} \\ \\
& = \int_{f^{-1}(\Omega)} d\vec{x} R(\vec{x})  e^{\kappa \tau(\vec{x})}  \rho_{c}(\vec{x})   \\ \\
& = \int_{\Omega} d\vec{x} R(\vec{x})  e^{\kappa \tau(\vec{x})}  \rho_{c}(\vec{x}) - \int_{E} d\vec{x} \rho_{c}(\vec{x}). 
\end{array}
\end{equation}
We used  $| \mathcal{D}_f(\vec{x}) |=| d\vec{x}'|/|d\vec{x}|$, and the fact that $f^{-1}(\Omega) \cap \Omega =\Omega \setminus E$. After rearrangement 
\begin{equation}
\ep=1+\mu_{c}(E), 
\label{eq.escaperate}
\end{equation}
where $\langle \ldots \rangle_c \equiv \int_\Omega \ldots d\mu_c= \int_\Omega \ldots \rho_c(\vec{x}) d\vec{x}$.
This new implicit formula for $\kappa$ involves the c-measure of map~(\ref{eq.extended}) and contains both $\tau(\vec{x})$ and 
$R(\vec{x})$.  It generalizes the Pianigiani-Yorke formula~\cite{PY:1979} $\kappa =-\ln[1-\mu_c(E)]$ valid for usual maps, for which  $\tau, R \equiv 1$
for $\vec{x} \in \Omega \setminus E$ while $\tau=0,R\equiv1$ for $\vec{x} \in E$. To see this, notice that (\ref{eq.escaperate}) can be
written as $\langle R e^{\kappa \tau} \rangle_c = e^\kappa[1-\mu_c(E)]+\mu_c(E)=1+\mu_c(E)$.

We now explore the implications of Eq.~(\ref{eq.escaperate}). As an approximation of a closed concert hall, consider the case of closed
systems ($E=\varnothing$) with homogeneous absorption  [$R(\vec{x})=R<1$] and nontrivial $\tau$'s, in which case~(\ref{eq.escaperate}) becomes $\langle
e^{\kappa \tau} \rangle_{c}=1/R$. Consider the cumulant expansion $\ln{\langle e^{\kappa \tau} 
  \rangle_{c}} = \sum_{r=1}^{\infty} (\kappa)^r C_r(\tau)/r!$, where $C_r$ are the cumulants of $\tau$ with respect to the c-measure.
 The $r=1$ approximant of $\kappa$ is $\kappa_1 = - \ln R / \langle \tau \rangle_c$. For $R\rightarrow 1$, we obtain $\langle \tau
 \rangle_c  \rightarrow \langle \tau \rangle$ and  $\kappa_1 \approx
(1-R)/\langle \tau \rangle$, which corresponds to Sabine's celebrated formula for the reverberation time~\cite{Kuttruff:2005},
where $\tcoll$ is the closed billiard average return time.
The $r=2$ approximant is
\begin{equation}\label{eq.kappa2}
\begin{array}{ll}
 \kappa_2 &= \frac{\sqrt{\tcoll^2_{c}-2 \sigma^2_{c}\ln{R}}-{\tcoll_{c}}}{\sigma^2_{c}}  \approx \kappa_1 \left( 1-\frac{\kappa_1}{2}\frac{\sigma^2_{c}}{\langle \tau \rangle_{c}} \right),
\end{array}
\end{equation}
where the approximation is valid for small variance~$\sigma_c$ of $\tau$ and was obtained in different contexts~\cite{Kuttruff:2005,Mortessagne:1992}.
The accuracy of these expressions depends on the rate of convergence of $\kappa_r
\rightarrow \kappa$, see Supplemental Material 
(SM) for details and general cases.

The importance of our general and exact formula~(\ref{eq.escaperate}) becomes clear in view of 
Joyce's pessimistic 
conclusion from 1975:
{\it It is further proven that the functional form of Sabine's expression cannot be modified so as to become correct for large
  absorption}~\cite{Joyce:1975}. While this negative result is an unavoidable consequence of 
  the argumentation being restricted to the properties of closed dynamics, Eq.~(\ref{eq.escaperate}) provides the answer to Joyce's
  search based on the modern theory of open dynamical systems~\cite{PY:1979,LaiTel-book}.

We now turn to the effect of $R$ and $\tau$ on the spectrum of fractal dimensions
$D_q$. In a closed system ($E=\emptyset$) trajectories visit the whole phase space and thus~$D_0$ equals the phase space dimension. We 
argue below that a nontrivial $D_q$ (multifractality) is obtained even in this case, and
that $D_q$ depends on both $R$ and $\tau$.
We illustrate this through four examples ({\bf I-IV}) with increasing complexity.


{\bf I.} Consider the tent map
$f(x)=ax$ with $a>2$ for $x<1/2$, and $f(x)=a(1-x)$ for $1/2 \le x \le 1$. We extend $f(x)$ by adding return times~$\tau(x)$ and reflection
coefficients~$R(x)$ which, for simplicity, are chosen to be constant on the two elements $i=1,2$ of the generating partition:
$(\tau_1, R_1)$ on 
$I_1=[0,1/a]$, and $(\tau_2, R_2)$
on $I_2=[1-1/a,1]$. 
The escape region is $E=(1/a<y<1-1/a)$ [where $(\tau,R)=(0,1)$]. 
Direct substitution into the steady state of (\ref{eq.FP2}) with $|\mathcal{D}_f|\equiv |f'|=a$ shows that $\rho_c=1$ on $x\in[0,1]$, and that the relation 
  for~$\kappa$ is
\begin{equation}\label{eq.unified}
P_1+P_2=1, \text{ with } P_i \equiv R_i e^{\kappa \tau_i}/a.
\end{equation}
To see that Eq.~(\ref{eq.unified}) is consistent with~(\ref{eq.escaperate}), notice that there are only three intervals ($I_1,I_2,$ and
  $E$) with different 
  $Re^{\kappa \tau}$ and, due to the constancy of $\rho_c$, their c-measure equal their length. It follows that
$\mu_{c}(E)=1-2/a$ and $\ep=R_1e^{\kappa \tau_1}/a+R_2e^{\kappa \tau_2}/a+(1-2/a)=2-2/a=1+\mu_c(E)$, where (\ref{eq.unified}) was used. 
$P_i$ in Eq.~(\ref{eq.unified}) can be interpreted as the proportion of weighted
trajectories, initiated uniformly in $I_i$, after one   
iteration of Eq.~(\ref{eq.FP2}). 
Analogously, the weights on the preimages of $I_i$ of length $1/a^2$ are $P_1^2$, $P_2 P_1$, $P_2^2$, and $P_1 P_2$. 
As will be clear from example {\bf II}, the continuation of this procedure provides a multifractal measure, $\mu$ different from
$\mu_c$, which corresponds to the weights on small intervals covering the never escaping points (chaotic repeller).

{\bf II.} Consider general non-invertible expanding maps $f(x)$ defined on $x\in[0,1]$ with general $\tau(x)$ and $R(x)$.  
In the most typical single humped family, the $n$-th preimages, of number $2^n$, of the unit interval are the so-called
cylinders~$I^{(n)}_i$~\cite{LaiTel-book}. 
In general, $\rho_c$ is not constant, but is continuous and covers $x\in[0,1]$.
A fractal measure, $\mu$, can be found by considering the analogues of the weights $P_1$ and $P_2$ for cylinders $P(I_i^{(n)})$. For shrinking cylinders, $P$ 
approximates the measure $\mu$ of the repeller $\mu_i^{(n)} \equiv \mu(I_i^{(n)}) \approx P(I_i^{(n)})$.
 To compute this, consider the $n$ fold iterated map $f^n$ and the corresponding
Eq.~(\ref{eq.FP2}).  
For large $n$ the logarithm of the slope of
$f^{n}$ at $x$ is approximately constant in a cylinder and therefore 
$P(I_i^{(n)})=e^{n \kappa \tau_i^{(n)}} e^{n \ln R_i^{(n)}} \epsilon_{i}^{(n)}$, where $\epsilon_{i}^{(n)} = 1/|(f^n)^{'}(x_0)|\equiv
e^{-\lambda_i^{(n)} n}$ with $x_0
\in I_i^{(n)}$. In turn, $\tau_i^{(n)}$ and $\ln R_i^{(n)}$ 
are sums of $\tau_{1,2}$ and $\ln{R_{1,2}}$, respectively,
over a typical trajectory $(x_0,...,x_j,...,x_{n-1})$ of length $n$ divided by $n$. 
Here $x_{n-1}$ is arbitrary, but fixed, and  $x_0 = f^{-n}(x_{n-1}) \in I_i^{(n)}$ for all cylinders.
Altogether,
\begin{equation}
\mu_i^{(n)} \sim e^{n (\kappa \tau_i^{(n)}+\ln R_i^{(n)}-\lambda_i^{(n)})}\sim \Pi_{j=0}^{n-1}\frac{e^{\kappa \tau(x_j)} R(x_j)}{\mid f'(x_j) \mid},
\label{mu}
\end{equation}
which hardly depends on $x_{n-1}$ (because of the shrinking cylinders) and differs from $\mu_c(I_i^{(n)})$ (which is proportional to $\epsilon_i^{(n)}$). 

Averages of an observable $A$ over the repeller measure $\mu$ are obtained as $\overline{A} \equiv
 \lim_{n\rightarrow\infty}\sum _{i=1}^{2^n} A^{(n)}_i \mu^{(n)}_i$ (e.g.,  the average Lyapunov exponent is $\bar{\lambda}_+ = \lim_{n\rightarrow\infty} \sum_i \lambda_i^{(n)} \mu_i^{(n)}$)
The information dimension of the repeller~$D_1$ follows from the general relation
$D_1 = \lim_{n\rightarrow\infty} {\sum_i \mu_i^{(n)} \ln{\mu_i^{(n)}} }/{\sum_i \mu_i^{(n)} \ln{\epsilon_i^{(n)}}}$ 
~\cite{Ott-book,LaiTel-book}.
Substituting (\ref{mu}), we find 
\begin{equation} \label{eq:ld_D1}
D_1 =1 - \frac{\kappa \bar{\tau} +\overline{\ln R}}{\bar{\lambda}_+}.
\end{equation}
Due to the {\it reflection rate} $\overline{\ln R}$, this is a generalization 
of the Kantz-Grassberger relation ($D_1=1-\kappa/\bar{\lambda}_+$)~\cite{KG:1985} to any chaotic 1D map with absorption.  
For the tent map of example {\bf I.},  $\lambda_+=\ln{a}$,  $\bar{\tau}=P_1 \tau_1 + P_2 \tau_2$, $\overline{\ln R}=P_1 \ln{R_1} + P_2
\ln{R_2}$, and the order-$q$ dimension can be calculated (from  $\sum_i \mu_i(\epsilon)^q \sim \epsilon^{(q-1)D_q}$
\cite{Ott-book}) as  
\begin{equation}\label{eq.Dq}
D_q=\frac{\ln{(P_1^q+P_2^q)}}{(1-q)\ln{a}}.
\end{equation}

\begin{figure}[!bt]
\centering
\includegraphics[width=0.8\columnwidth]{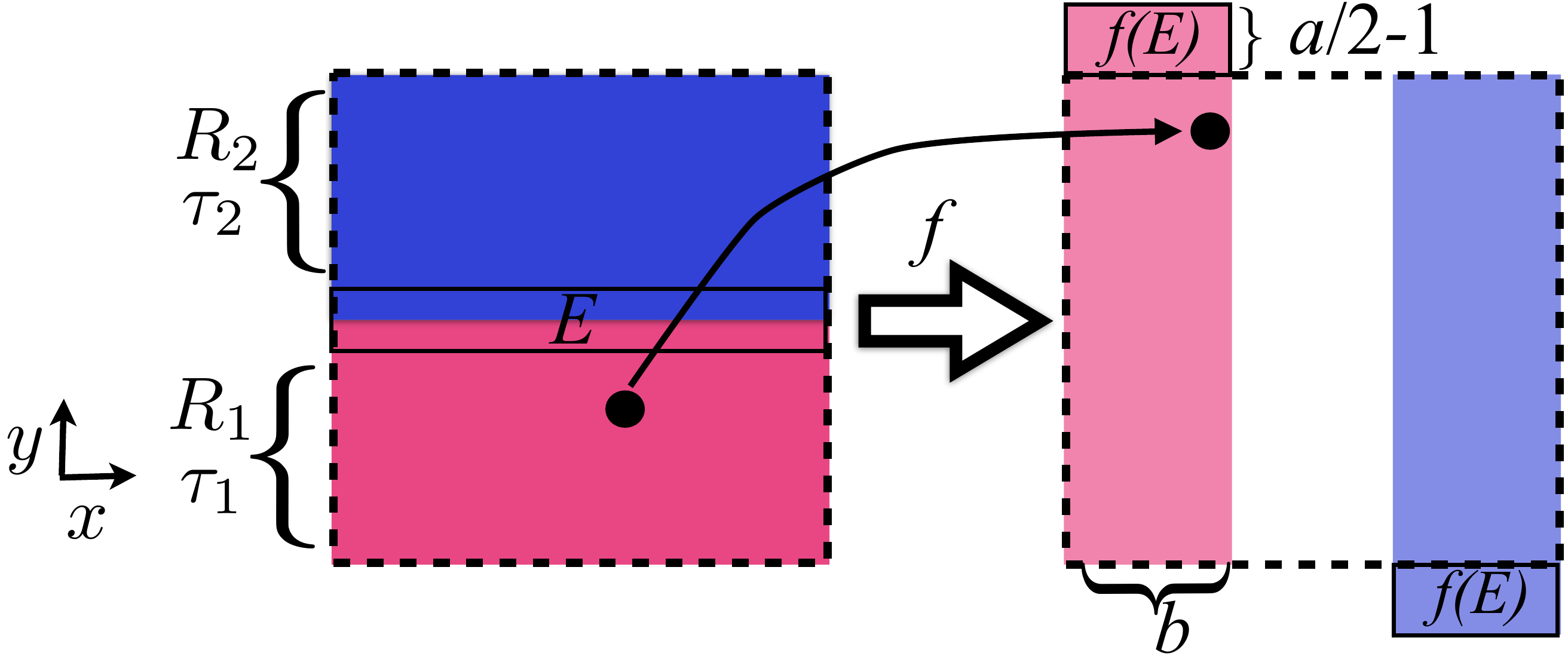}
\caption{
Open baker map with absorption and return times. Intensity~$J$ decays due to $R<1$. Trajectories leave the system
(unit square) when $(x,y) \in E$. The extended map~(\ref{eq.extended})
  is $(x',y') = (bx, ay)  \text{ for } y<1/2$, $(x',y')= 1-b(1-x),1-a(1-y)$ for $y\ge1/2$,
  $(\tau,R)=(\tau_1,R_1)$ if  $y<1/a$ and $(\tau_2,R_2)$ if  $y>1-1/a$. $b\le 1/2$, $a\ge 2$; $\mathcal{D}_f=ab$. 
}
\label{fig.baker}
\end{figure}

{\bf III.} We now apply our operator formalism~(\ref{eq.FP2}) to an invertible 2D map, the analytically solvable 
baker map, see
Fig.~\ref{fig.baker}~\cite{Ott-book,LaiTel-book}. Consider initially $\rho_0 \equiv 1$. In the next step, $\rho_0$ is multiplied
by $R_i e^{(\kappa \tau_i)}/(ab)$, $i=1$ or $2$, leading to two columns of width $b$ parallel to the $x=0$  and $x=1$ axes with measures $P_1$
and $P_2$, respectively, as given in (\ref{eq.unified}).
The construction goes on in a self-similar way. 
Prescribing that the c-measure corresponds to a case when
the full measure $(P_1+P_2)^n$ after $n \gg 1$ steps remains unity, Eq.
(\ref{eq.unified}) is recovered (the dynamics along unstable manifolds of 2D maps is faithfully represented by 1D maps). In addition, $P_i$'s are the c-measures 
of the columns of width $b$ and of unit height.

Concerning the dimensions of the c-measure~$D_{q,c}$,  
we concentrate on the partial dimensions $D_q^{(2)}$ along the stable ($x$) direction because $\mu_c$ is constant along $y$ and therefore $D_{q,c}=1+D_q^{(2)}$.
After $n$ steps, the boxes
in the  $x$-direction are of length $b^n$ and therefore
$D_q^{(2)}=D_q \lambda_+/\lambda_-$, where $\lambda_-=-\ln{b}$
is the modulus of the contracting Lyapunov exponent and $D_q$ is given by (\ref{eq.Dq}).
Although~(\ref{eq.Dq}) was obtained as $D_q$ of the tent map repeller, an analogous procedure applied to the horizontal bands of height $(1/a)^{n}$ yields that the order-$q$ dimension of the baker saddle's stable 
manifold is $1+D_q$~\footnote{This stable manifold measure is consistent with the usual definition~\cite{Ott-book}.}.  
$D_q$ can therefore be considered to be the partial dimension 
$D_q^{(1)}$ along the unstable $(y)$ direction $D_q^{(1)}=D_q$.
The
dimension of the saddle is  $D_q^{(1)}+D_q^{(2)}$.

As an example, consider the
closed area preserving map ($a=1/b=2$) with weak absorption $1-R_i \ll 1$, for which $\kappa$ is small. 
Assuming $\kappa$ to be of the same order as $1-R_i$, in leading order, $P_i=(1-(1-R_i)
+\kappa \tau_i)/2$ and, from (\ref{eq.unified}), $\kappa=(1-R_1+1-R_2)/(\tau_1+\tau_2)$. 
Inserting $P_1=(1-\Delta)/2$ and $P_2=(1+\Delta)/2$ into (\ref{eq.Dq}), we obtain  
\begin{equation}
D_q^{(1)}=1-q \frac{\Delta^2 }{2 \ln2} \;\; \text{ and } \;\; D_{q,c}=2-q \frac{\Delta^2 }{2 \ln2},
\label{eq.Dq0}
\end{equation}
valid for $0\le q< 1/{\Delta}^2$, where $\Delta=[(1-R_1)\tau_2 - (1-R_2)\tau_1]/(\tau_1+\tau_2)$, and
$D_q^{(2)}=D_q^{(1)}$ because $\lambda_+=\lambda_-$.
This illustrates that both inhomogeneous absorption ($R_1 \neq R_2$) and return time ($\tau_1 \neq \tau_2$) distributions
lead to $D^{(i)}_q\neq D^{(i)}_0=1$. Multifractality becomes {\it stronger} 
with increasing absorption. In contrast, for the usual closed area preserving baker map $\kappa=0$ and $D_q^{(1,2)}=1$, illustrating how the
results from the traditional operator~(\ref{eq.FP1}) and the generalized operator~(\ref{eq.FP2}) can differ even for the same map~$f$.

\begin{figure}[!bt]
\includegraphics[width=1\columnwidth]{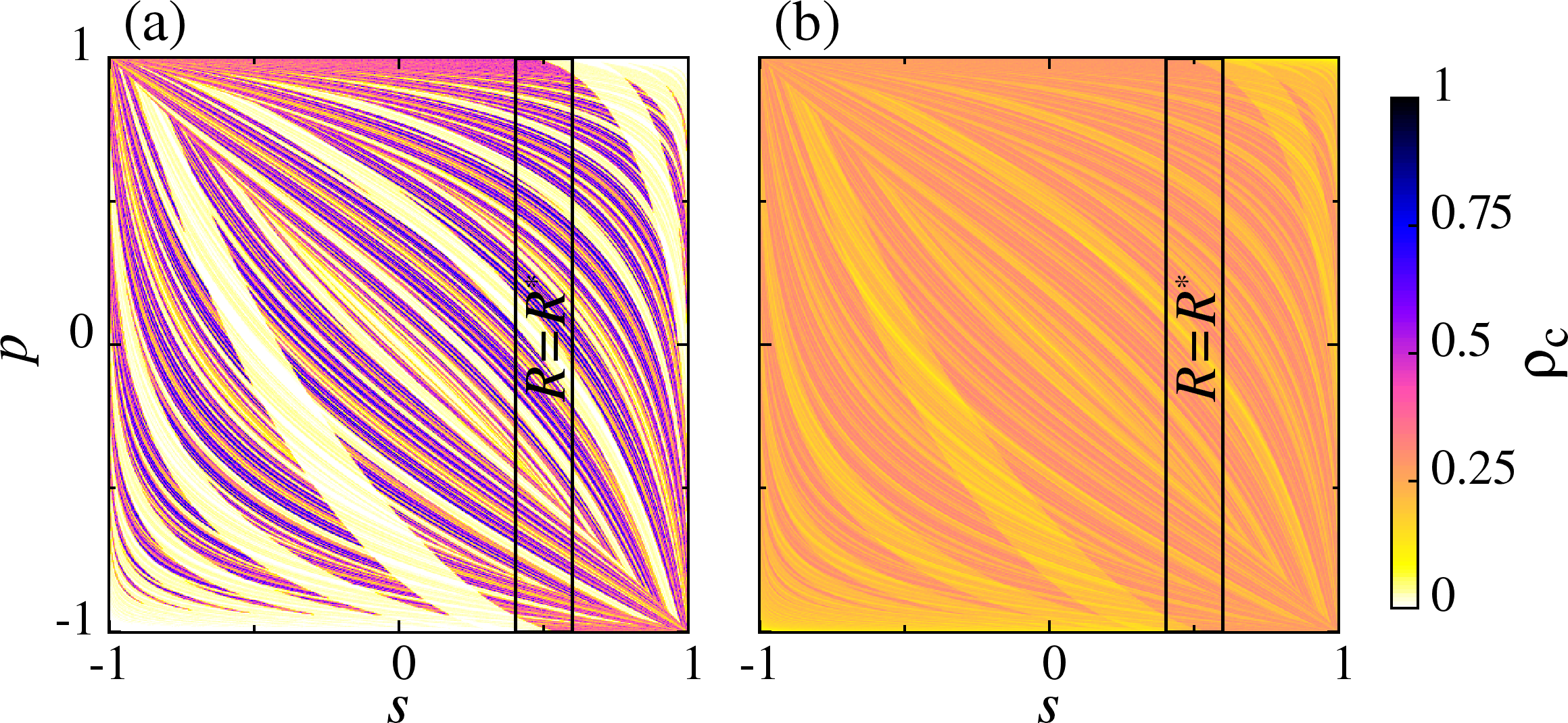}
\caption{Conditionally invariant density~$\rho_c$ for the cardioid billiard with absorption. As shown in Fig.~\ref{fig1}, $R=1$ everywhere except in $\{ s \in [0.4,0.6], p\in [-1,1]\}$ where $R=R^*<1$. (a) $R^*=0.1$; and (b) $R^*=0.75$. 
Structures in (b) amount to $0.2\%$ difference between $D_{0,c}$ and $D_{1,c}$, see Tab.~\ref{tab.cardioid}.}
\label{fig3}
\end{figure}

{\bf IV.} Our final example is the fully chaotic cardioid billiard with an absorbing segment of the boundary where $R=R^*<1$, see Fig.~\ref{fig1}~\cite{Robnik1983}.
 Figure~\ref{fig3} shows $\rho_c$ for two values of $R^*$, computed using ray simulations (\ref{eq.extended})~\cite{RMP}. $D_{q,c}$ for $q=0, 1,$ and $10$
are reported in Tab.~\ref{tab.cardioid} and exhibit 
$R$-dependent multifractality like in the baker map. A comparison with the $R^*=0$ case (trajectories escape) shows that
the slightest nonzero $R^*$ without trajectory escape leads to a space-filling unstable manifold ($D_{0,c}=2$) whose $D_{1,c}$
is close to $D_{0,c}$ of the $R^*=0$ case. The difference between $D_{0,c}$ and $D_{1,c}$ quantifies the enhancement in multifractality due to absorption.

{
\renewcommand{\tabcolsep}{3pt}
\begin{table}[bt]
\begin{tabular}{|c | c c c c c @{$\quad$} | c|}
\hline
\multicolumn{1}{ |c| }{$R^*$} & $0.01$ & $0.05$ & $0.25$ & $0.5$ & $ 0.75$ & $0$ \\
\hline
\hline
$D_{0,c}$ & $2.00$  & 2.00 & 2.00 & 2.00 & 2.00 & 1.84 \\
$D_{1,c}$ & 1.84 & 1.87 & 1.94 & 1.981 & 1.996 & 1.82 \\
$D_{10,c}$ & 1.79 & 1.80 & 1.86 & 1.923 & 1.975 & 1.75 \\
\hline
$D_{1,c}^{Eq. (\ref{eq.twelve})}$ & 1.83 & 1.86 & 1.94 & 1.980 & 1.996 & 1.81 \\
\hline
\hline
$\kappa$ & 0.06470 &	0.06155 &	0.04663 &	0.02954 &	0.01410 & 0.06559	\\
\hline
$\kappa'_1$ & $0.06434$ & $0.06131	$ & $0.04641	$ & $0.02945$ & $0.01408$ & $0.06520$\\
$\kappa'_2$ & $0.06468$ & $0.06163	$ & $0.04660	$ & $0.02953$ & $0.01410$ & $0.06557$\\
\hline
\end{tabular}
\caption{ Escape rate~$\kappa$ and order-q dimensions $D_{q,c}$ of the c-measure of the
  cardioid billiard described in Figs.~\ref{fig1} and \ref{fig3} for different $R^*$. 
$D_{q,c}$ is measured from the c-measure of partitions of the phase space (Fig.~\ref{fig3}) and
$D_{1,c}^{Eq. (\ref{eq.twelve})}$ from the saddle's measure and Eq.~(\ref{eq.twelve})
   (we found that $\bar{\lambda}_+=\bar{\lambda}_- \approx 0.35 \bar{\tau}$ for all $R^*$,
   see SM Fig.~S1).
$\kappa$ is measured by fitting the survival probability and $\kappa'_r$ 
are approximants obtained
directly from $\rho_c$, see SM (text and Figs.~S2 and S3).
} \label{tab.cardioid}
\end{table}
}

In summary, we argued that chaotic systems with absorption should be considered as a class of dynamical systems on its own. Absorption
converts the closed dynamics of trajectories into an open dynamics of weighted rays which we have shown to have 
fundamentally different chaos characteristics when compared to those of
traditional open systems (in which trajectories escape). Among such properties are the new Perron-Frobenius operator~(\ref{eq.FP2}), an implicit formula~(\ref{eq.escaperate}) for the escape rate, a generalized Kantz-Grassberger
relationship~(\ref{eq:ld_D1}) for $1$D maps, and an enhanced multifractality of 
invariant measures. 
We anticipate that absorption has also important consequences in other operator approaches based on Markov partitions, which received
renewed interest with the concept of almost invariant sets~\cite{Froyland:2010} and Ulam's method~\cite{Cristadoro:2013}. Furthermore, we conjecture that, provided the 
direct product structure seen in the baker example holds, for invertible chaotic 2D maps with absorption 
\begin{equation}\label{eq.twelve}
D_{1,c}=1+\left(1 - \frac{\kappa \bar{\tau}+\overline{\ln
  R}}{\bar{\lambda}_+}\right)\frac{\bar{\lambda}_+}{\bar{\lambda}_-}
\end{equation}
(see Tab.~\ref{tab.cardioid} for a numerical test).
Our results apply to
 any chaotic system with absorption or partial reflection, provide new relations that have been looked after for decades~\cite{Joyce:1975},
 have direct implications for wave-chaotic systems~\cite{Wiersig2008,Lu2003,Nonnenmacher:2008}, and are directly accessible to experiments
 (e.g., measuring the spatial distribution of decaying states in optical and acoustic systems).

\acknowledgments
We are indebted to G. Dr\'otos, H. Kantz,  Z. Kaufmann, R. Klages, P. Grassberger, and T. Weich for useful discussions. 
This work was supported by 
OTKA grant No. NK100296, the von Humboldt Foundation, and the Fraunhofer Society.

\newpage
\begin{widetext}
$\;$
\newpage
\section{Supplemental Material}

\section{Complete cumulant expansion of $\kappa$}

Here we evaluate different approximants from
the escape rate formula -- Eq. (5) in the main text -- for the general case of non-constant reflection
coefficient $R$ and non-zero exit region $E$. We write the main formula as
\begin{equation}\label{eq.formulaE}
\langle R e^{\kappa \tau} \rangle_c=\langle e^{\ln{R}} e^{\kappa \tau} \rangle_c=1+\mu_c(E),
\end{equation}
and the corresponding cumulant expansion is
\begin{equation}\label{eq.cumdef}
\ln{\langle e^{s}\rangle_c}=\sum_{r=1}^\infty C_r(s)/r!=\ln[1+\mu_c(E)],
\end{equation}
where $s=\kappa \tau +\ln{R}$. Up to second order in $r$, Eq.~(\ref{eq.cumdef}) yields
\begin{equation}
\kappa \langle\tau\rangle_c + \langle \ln{R} \rangle_c + \frac{1}{2} \sigma_s^2=\ln[1+\mu_c(E)]. 
\label{eq.cum}
\end{equation} 
Here $\sigma_s^2$ represents the second cumulant of variable $s$.
By definition
\begin{equation}
\sigma_s^2=\langle(\kappa \tau +\ln{R})^2\rangle_c -(\kappa \langle\tau\rangle_c +\langle\ln{R}\rangle_c)^2. 
\end{equation}  
Writing out the terms explicitly,
\begin{equation}
\sigma_s^2=\kappa^2(\langle\tau^2\rangle_c -\langle\tau\rangle_c^2)+2 \kappa (\langle\tau \ln{R}\rangle_c - \langle\tau\rangle_c\langle\ln{R}\rangle_c) +
\langle(\ln{R})^2\rangle_c-\langle\ln{R}\rangle_c^2 \equiv \kappa^2 \sigma_{\tau}^2+2 \kappa \alpha +\sigma_{\ln{R}}^2,
\end{equation}
where the last line defines the second cumulants $\sigma_{\ln{R}}$,  $\sigma_\tau$ ($\equiv\sigma_c$ of the main text), and $\alpha$ (the latter being more 
a cross correlation than a second cumulant). 
Altogether we obtain a quadratic equation
\begin{equation}
\frac{1}{2} \kappa^2 \sigma_{\tau}^2 + \kappa (\langle\tau\rangle_c +\alpha)+ \langle\ln{R}\rangle_c  +\sigma_{\ln{R}}^2/2 -\ln[1+\mu_c(E)]= 0.
\label{eq.quadE}
\end{equation}
The first ($r=1$) order approximant $\kappa_1$ follows by neglecting  $\sigma_s^2$ in Eq.~(\ref{eq.cum}) 
\begin{equation}
\kappa_1= -\frac{\langle\ln{R}\rangle_c-\ln[1+\mu_c(E)]}{\langle\tau\rangle_c},
\label{k1E}
\end{equation}
and $\kappa_2$  is obtained directly from~(\ref{eq.quadE}) as
\begin{equation}
\kappa_2=\left[ -(\langle\tau\rangle_c +\alpha) +\sqrt{(\langle\tau\rangle_c +\alpha)^2-2 \sigma_{\tau}^2 [\langle\ln{R}\rangle_c
    +\sigma_{\ln{R}}^2/2 -\ln(1+\mu_c(E))] }\right] / \sigma_{\tau}^2.
\label{k2E}
\end{equation}
The solution with minus sign is not relevant
since it does not recover (\ref{k1E}) for vanishing second cumulants. The approximants
$\kappa_1$ and $\kappa_2$ in Eqs.~(\ref{k1E}) and~(\ref{k2E}) are the most general forms
(non-constant $R$ and non-empty $E$) of the
expressions for $\kappa_1$ and $\kappa_2$ in the main text. 

A fast (in $r$) convergence of the series $\kappa_r$ to $\kappa$ is assured in two cases: small values of $\kappa$ and small cumulants 
of the variable $s=\kappa \tau +\ln{R}$. Therefore, for large fluctuations of $R$ in
the phase space we expect, 
e.g., $\kappa_2$ to be
far from $\kappa$. This situation is particularly important if regions with $R=1$ and $R\gtrapprox 0$ are present
simultaneously in the phase space (as in the Cardioid example with $R^*\ll 1$ considered in the last part of the manuscript). This motivates
us to consider an alternative expansion.

\section{Alternative cumulant expansion of $\kappa$}

In a previous publication~\cite{RMP}, we have considered an alternative convention for the return time $\tau$. Instead of attributing
$\tau$ to the initial point $\vec{x}$, as considered above and in the main text, in Ref.~\cite{RMP} we considered invertible maps and
attributed $\tau$ to the 
end point $\vec{x}'$ between two intersections of the Poincar\'e surface of section $\vec{x}'=f(\vec{x})$. We denote the return time 
$\tau(\vec{x}')$ 
as $\tau'$. With this convention, it is natural to move
the $e^{\kappa \tau}$ term as $e^{-\kappa \tau'}$ to the left hand side of our generalized operator formalism, Eq. (3) of the main
paper. In Ref.~\cite{RMP}, using the same reasoning as in the main text,
instead of Eq.~(\ref{eq.formulaE}), we
obtained the  following implicit formula for invertible systems
\begin{equation}\label{eq.formulaE2}
\langle e^{-\kappa \tau'} \rangle_c= \langle R \rangle_c,
\end{equation}
in which case the exit region $E$ corresponds simply to a region where $R=0$ 
(in the previous convention this was not the case because of the
choice $R=1$, $\tau=0$ for $\vec{x} \in E$). Using the return time convention consistently, the two descriptions are
equivalent so that $\kappa$ appearing in (\ref{eq.formulaE}) and (\ref{eq.formulaE2}) are the same. However, the approximants from
the cumulant expansions typically do not coincide. Now, the relevant variable 
is 
$s'=-\kappa \tau'$.  We denote the approximants obtained from~(\ref{eq.formulaE2}) by $\kappa'_r$ and write
similarly to~(\ref{eq.cumdef})
\begin{equation}\label{eq.cumdef2}
\ln{\langle e^{s'}\rangle_c}=\sum_{r=1}^\infty C_r(s')/r!=\ln \langle R \rangle_c.
\end{equation}
Up to second order in $r$ we obtain
\begin{equation}\label{eq.cum2}
-\kappa \langle\tau'\rangle_c + \frac{\kappa^2}{2} \sigma^2_{\tau'}=\ln \langle R \rangle_c,
\end{equation} 
which is simpler than Eqs.~(\ref{eq.cum}) and~(\ref{eq.quadE}) obtained in the previous convention. The first ($r=1$) order approximant
$\kappa'_1$ follows by neglecting $\sigma_{\tau'}^2$ as

\begin{equation}
\kappa'_1= -\frac{\ln{\langle R\rangle_c}}{\langle\tau'\rangle_c},
\label{k1E2}
\end{equation}
and $\kappa'_2$  is obtained directly from~(\ref{eq.cum2}) as
\begin{equation}\label{k2E2}
\begin{array}{ll}
 \kappa'_2 &= \left[{\langle \tau' \rangle_c}-\sqrt{\tau'^2_c+2 \sigma^2_{\tau'} \ln \langle R \rangle_c} \right]/{\sigma^2_{\tau'}} \\       
       & = \kappa'_1 \left( 1+\frac{\kappa'_1}{2}\frac{\sigma^2_{\tau'}}{\langle \tau' \rangle_c}+...\right).
\end{array}
\end{equation}
The last equation resembles Eq.~(6) in the main text, but differs from it due to the minus sign and due to the different return time convention.

The advantage of the approximants $\kappa'_{1,2}$ in Eqs.~(\ref{k1E2}) and~(\ref{k2E2}) over $\kappa_{1,2}$ in Eqs.~(\ref{k1E})
and~(\ref{k2E}) is that the reflection coefficient $R$ is not part of  variable $s'$ used in the cumulant expansion. Therefore, the
dependence on $R$ in Eqs.~(\ref{k1E2}) and~(\ref{k2E2}) appears only in the term $\ln \langle R \rangle_c$ and not in higher cumulants of
$ \ln R$ as in Eqs.~(\ref{k1E}) and~(\ref{k2E}). This difference is expected to be crucial for cases with strongly non-uniform $R$. 
Indeed, while $\kappa'_{1,2}$ are in excellent agreement with all numerical simulations in the Cardioid billiard (see Tab. I,
main text), we observed that $\kappa_{1,2}$ agreed very well with the numerically
determined value of $\kappa$ when $R^*$ was close to unity,
 but not for the cases with small $R^*$.  Both $\kappa'_{1,2}$ and $\kappa_{1,2}$ agree well  for open cases when trajectories escape, which
 is achieved in Eq.~(\ref{eq.cumdef}) taking uniform $R=1$ and including an exit set $E$ or, alternatively, in Eq.~(\ref{eq.cumdef2}) taking
 $R=1$ everywhere except in the interval corresponding to the exit set E of the previous case, where $R^*=0$. 
Beyond their conceptual relevance, the approximants $\kappa_{1,2}$ and $\kappa'_{1,2}$ are of practical use whenever the c-density $\rho_c(x)$ can be efficiently estimated independent of trajectory based simulations (notice that $\rho_c(x)$ is needed to compute the averages $\langle...\rangle_c$ and all cumulants in all equations above).

\begin{figure}[!ht]
\includegraphics[width=0.7\columnwidth]{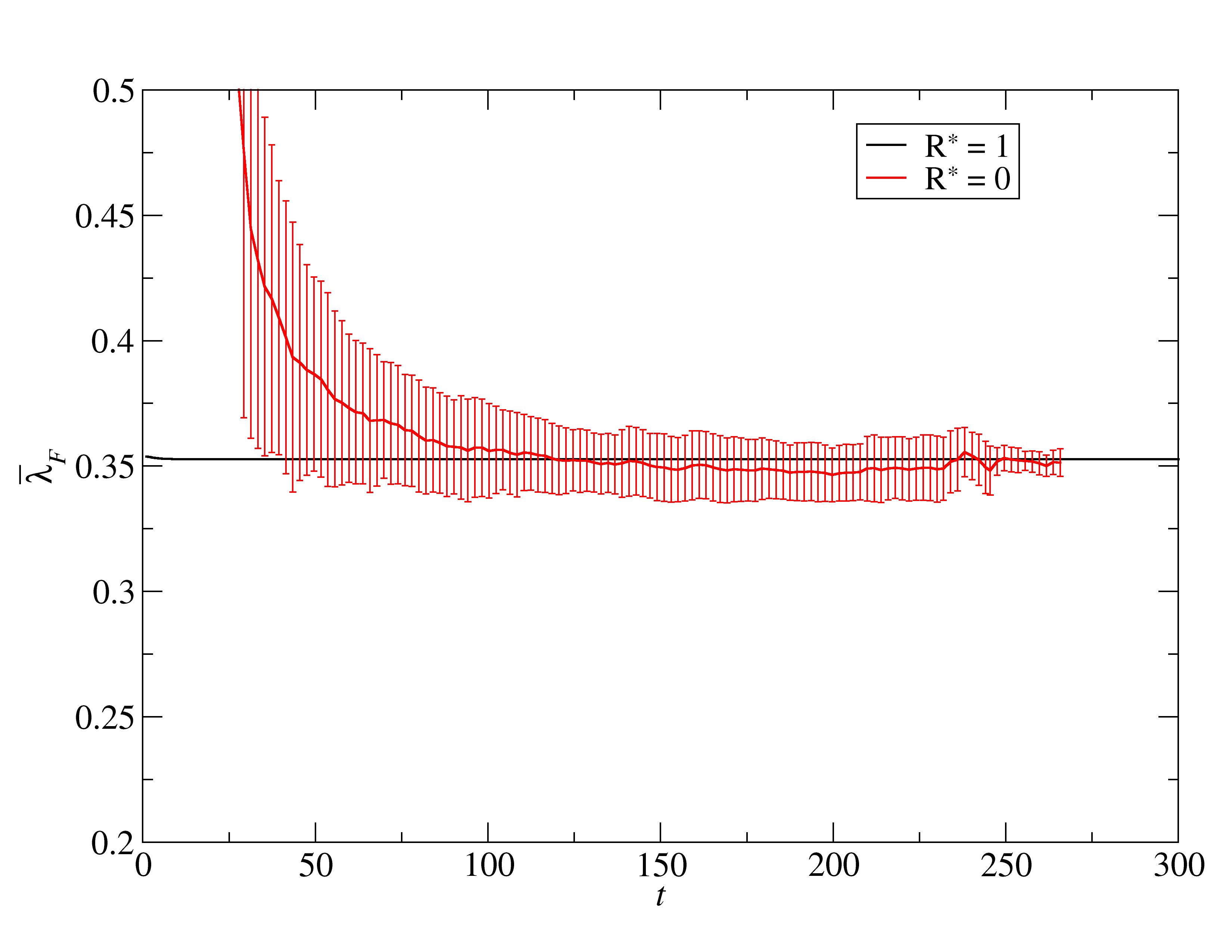}
\caption{Numerical computation of the flow Lyapunov exponent $\bar{\lambda}_F = \bar{\lambda}_+
  / \bar{\tau}$ for the Cardioid billiard as in
  Tab. I of the manuscript. At each time $t$, $\bar{\lambda}_F$  was 
  computed over trajectories that survive at least until time $t+t^*$. For $t\gg t^*\equiv
  1/\kappa ~ (\approx 15$ for the $R^*=0$ case) these trajectories at time $t$ provide
  good approximations of the chaotic saddle's measure. The reported value and error bars
  were estimated as the mean and standard deviation over $14$ different trajectories for
  $R=0$ and $90,000$ for $R=1$. For $0<R^*<1$ the estimations of $\bar{\lambda}_F$ fall in
  between the $R^*=0$ and the $R^*=1$ curves. Altogether, these results indicate that
  there is no significant variation of $\bar{\lambda}_F$ with $R^*$.} 
\end{figure}

\begin{figure}[!ht]
\includegraphics[width=0.7\columnwidth]{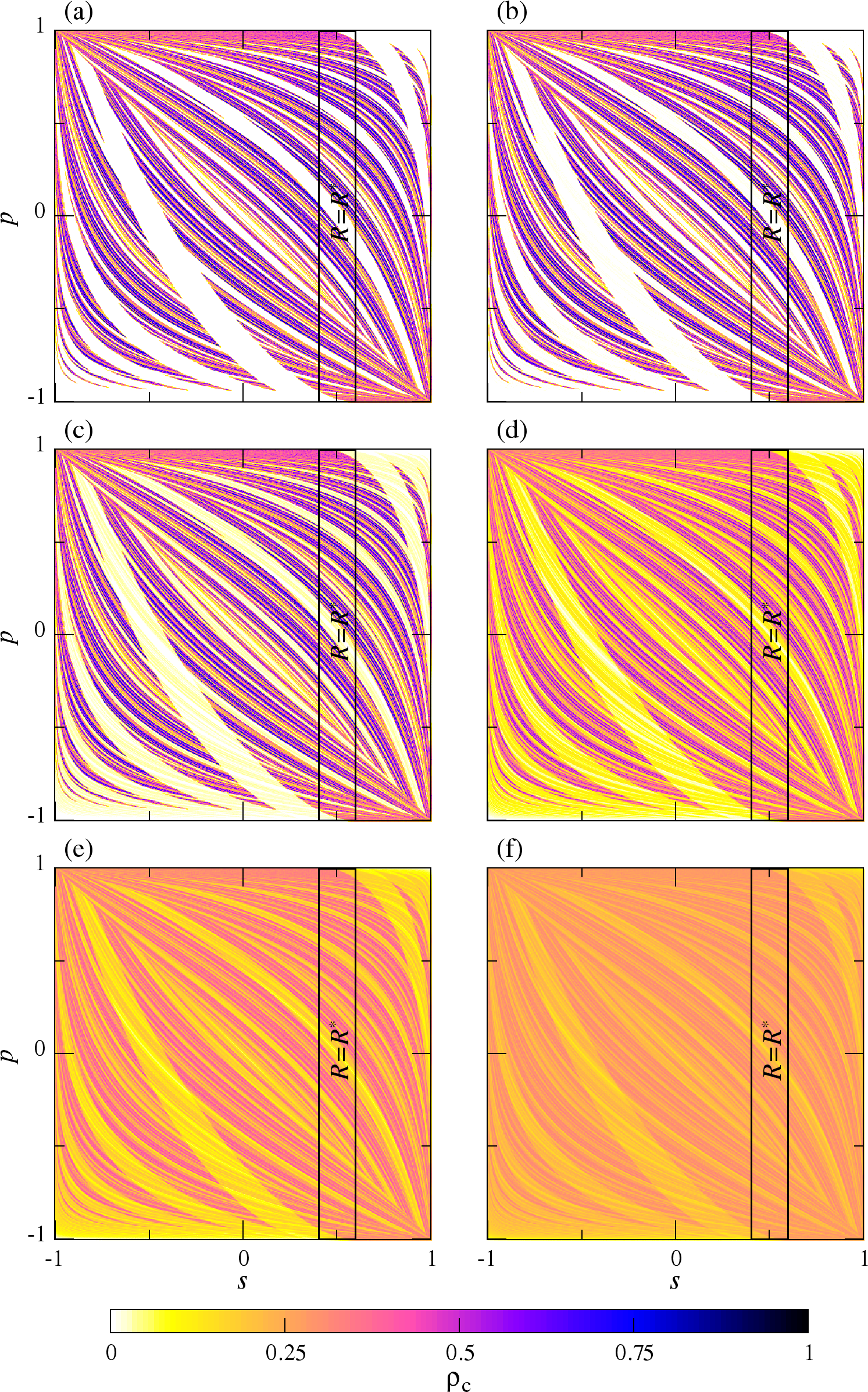}
\caption{Conditionally invariant density~$\rho_c$ for the cardioid billiard with absorption. The reflectivity is $R=1$ in the full perimeter
  except for the marked region $\{ s \in [0.4,0.6], p\in [-1,1]\}$ for which $R=R^*<1$, see Figs. 1 and 3 in the main text. (a) $R^*=0$; (b)
  $R^*=0.01$; (c) $R^*=0.05$; (d) $R^*=0.25$; (e) $R^*=0.5$; and (f) $R^*=0.75$. The sequence of plots of $\rho_c$ for increasing 
  values of $R^*$ evidences a smooth dependence on $R^*>0$. At $R^*=0.75$, $\rho_c$ tends to the uniform distribution $\rho_c=0.25$ (as the
  phase-space area is 4) of the fully closed billiard, keeping nonetheless the same structure, albeit at vanishing amplitude differences.
This structure is essentially the unstable manifold of the system fully opened in the marked region (R*=0), i.e. of the case when trajectories do escape.
 }
\label{figsm1}
\end{figure}

\begin{figure}[!ht]
\includegraphics[width=0.7\columnwidth]{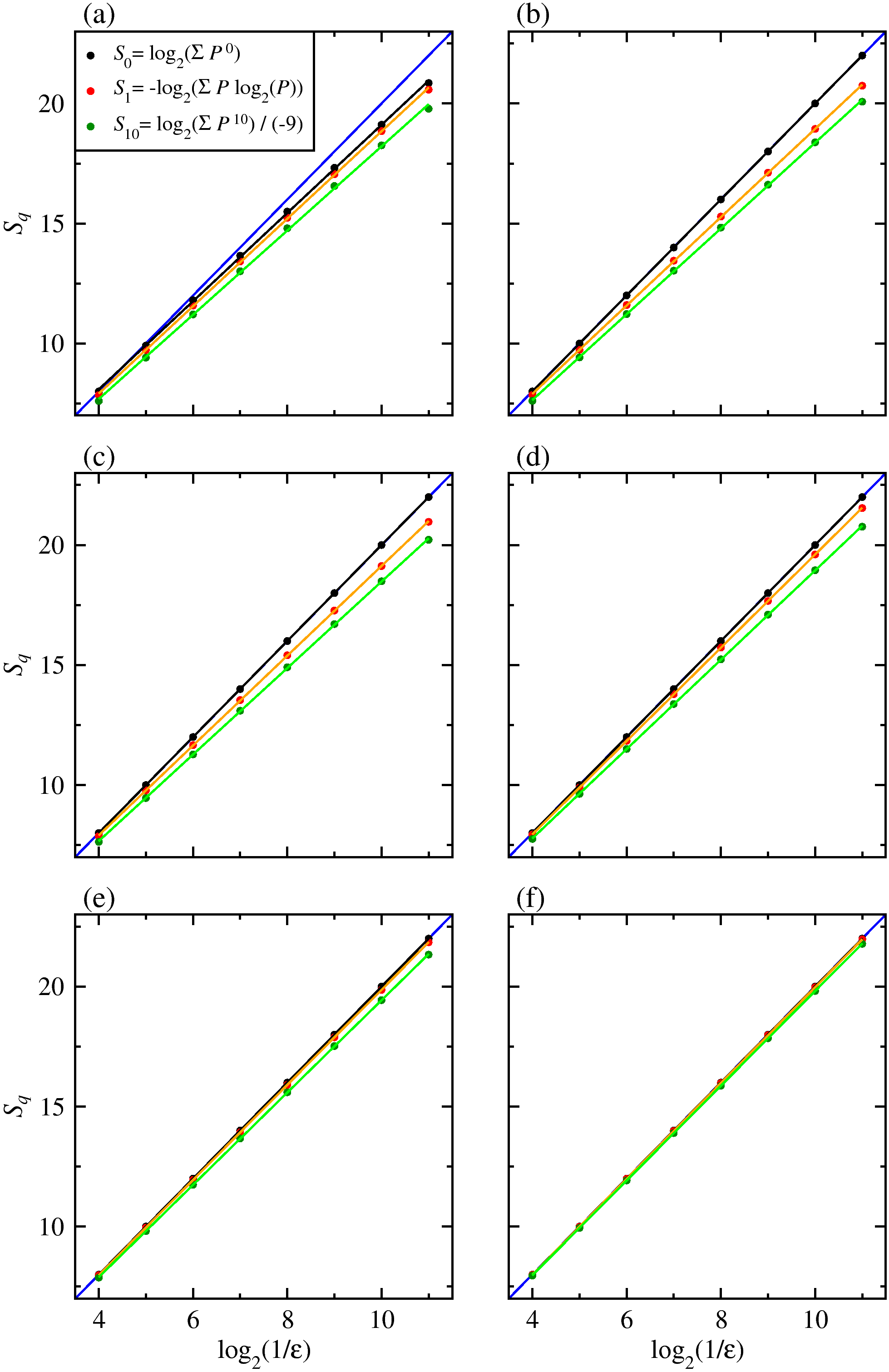}
\caption{Determination of order-$q$ fractal dimensions $D_{q,c}$ of the $c$-measure of the cardioid billiard for different values of the
  reflection coefficient $R^*$ (see Tab.~1, main text, and Fig.~\ref{figsm1}). (a) $R^*=0$; (b) $R^*=0.01$; (c) $R^*=0.05$; (d) $R^*=0.25$; (e) $R^*=0.5$; and (f) $R^*=0.75$. The estimates were computed directly from the
  definition, $D_{q,c}=S_q/\log_2(1/\epsilon)$, where $\epsilon$ is the 
  box edge length, $S_q=\log_2(\sum_i P_i^q)/(1-q)$, and $P_i$
  is the weight (c-measure) of the $i$-th box. The $P_i$ were obtained from approximations of the $c$-measure containing $\sim\!10^9$
  points. The reference blue line corresponds to a $D_{q,c}=2$ scaling and the remaining lines fit linearly the $S_q(\epsilon)$ obtained
  numerically (circles).  Fitting is restricted to a range where boxes are still well populated (up to $\sim\!1000$/box on
  average). The uncertainty of the fitting was estimated considering the fluctuations of the slope of neighboring points and is of the order of 
    the last reported digit in Tab.~1, main text.} 
\end{figure}

\end{widetext}

\end{document}